\documentclass{PoS}

\usepackage{amsmath}
\usepackage{amssymb}
\usepackage{graphicx}  

\newcommand{\beq}{\begin{equation}}
\newcommand{\eeq}{\end{equation}}
\newcommand{\bea}{\begin{eqnarray}}
\newcommand{\eea}{\end{eqnarray}}

\title{Flux tubes in the SU(3) vacuum}

\ShortTitle{Contribution title}

\author{Mario Salvatore Cardaci\\
        Dipartimento di Fisica dell'Universit\`a della Calabria \\
        and INFN - Gruppo collegato di Cosenza, 
        I-87036 Arcavacata di Rende, Cosenza, Italy\\
        E-mail: \email{salvatore.cardaci@fis.unical.it}}

\author{Paolo Cea\\
        Dipartimento di Fisica dell'Universit\`a di Bari \\ 
        and INFN - Sezione di Bari, I-70126 Bari, Italy \\
        E-mail: \email{paolo.cea@ba.infn.it}}

\author{Leonardo Cosmai\\
        INFN - Sezione di Bari, I-70126 Bari, Italy \\
        E-mail: \email{leonardo.cosmai@ba.infn.it}}

\author{Rossella Falcone\\
        Fakult\"{a}t f\"{u}r Physik, Universit\"{a}t 
        Bielefeld, Postfach 100131, D-33615 Bielefeld, Germany \\
        E-mail: \email{rfalcone@physik.uni-bielefeld.de}}

\author{\speaker{Alessandro Papa}\\
        Dipartimento di Fisica dell'Universit\`a della Calabria \\
        and INFN - Gruppo collegato di Cosenza, 
        I-87036 Arcavacata di Rende, Cosenza, Italy\\
        E-mail: \email{papa@cs.infn.it}}

\abstract{We analyze the distribution of the chromoelectric field generated by
a static quark-antiquark pair in the SU(3) vacuum. We find that the
transverse profile of the flux tube resembles the dual version of the
Abrikosov vortex field distribution and give an estimate of the London
penetration length in the confined vacuum.}

\FullConference{ The XXIX International Symposium on Lattice Field Theory - Lattice 2011\\
July 10-16, 2011\\
Squaw Valley, Lake Tahoe, California}

\begin{document}

\section{Introduction}

Color confinement in Quantum Chromo-Dynamics (QCD) is a long-distance 
behavior whose understanding continues to be a challenge for theoretical 
physics~\cite{Bander:1980mu,Greensite:2003bk}. Tube-like structures emerge by 
analyzing the chromoelectric field between static 
quarks~\cite{Fukugita:1983du,Kiskis:1984ru,Flower:1985gs,
Wosiek:1987kx,DiGiacomo:1989yp,DiGiacomo:1990hc,Singh:1993jj,Cea:1992sd,
Matsubara:1993nq,Cea:1992vx,Cea:1993pi,Cea:1994ed,Cea:1994aj,Cea:1995zt,
Bali:1994de,Haymaker:2005py,D'Alessandro:2006ug}. 
Such tube-like structures naturally lead to linear potential and 
consequently to a ``phenomenological'' understanding of color confinement.
To explain the formation of chromoelectric flux tubes in QCD vacuum, 
't Hooft~\cite{'tHooft:1976ep} and Mandelstam~\cite{Mandelstam:1974pi}
proposed the hypothesis that QCD vacuum behaves like a coherent state of color 
magnetic monopoles, which leads to the picture of QCD vacuum a magnetic 
(dual) superconductor~\cite{Ripka:2003vv}. 
According to this picture the observed color flux tubes are naturally accounted
for by the (dual) Meissner effect, in analogy with the formation of 
Abrikosov tubes in the usual superconductivity~\cite{Abrikosov:1957aa}.
Even if the 't Hooft construction does not explain the dynamical formation of 
color magnetic monopoles, many lattice calculations~\cite{Shiba:1994db,
Arasaki:1996sm,Cea:2000zr,Cea:2001an,DiGiacomo:1999fa,DiGiacomo:1999fb,
Carmona:2001ja,Cea:2004ux,D'Alessandro:2010xg} have given numerical evidence 
in favor of magnetic monopole condensation in the QCD vacuum. 

On the other side, magnetic monopole condensation could be the 
consequence rather than the origin of the confinement 
mechanism~\cite{'tHooft:2004th}.
Even in this case, however, the dual superconductivity picture provides us 
with a ``phenomenological'' frame for the analysis of tube-like structure in 
the QCD vacuum.

The outcome of previous studies~\cite{Cea:1992vx,Cea:1993pi,Cea:1994ed,
Cea:1994aj,Cea:1995zt} of the SU(2) confining vacuum was the following:
(1) presence in lattice configurations of color flux tubes made up by 
the chromoelectric fields directed along the line joining a static 
quark-antiquark pair; (2) transverse size of the chromoelectric flux tube 
interpreted as the London penetration length in the Meissner effect;
(3) penetration length measured both in the maximal Abelian gauge and 
without gauge fixing, with compatible results, thus supporting 
gauge-invariance; (4) determination of the string tension as the energy 
stored into the flux tube per unit length, in good agreement with the results 
in the literature.

The aim of the present work is to extend the analysis to the more interesting 
case of the SU(3) gauge theory.

\section{Color fields on the lattice}
\label{numerical}

\begin{figure}[tb]
\centering
\includegraphics[width=0.40\textwidth]{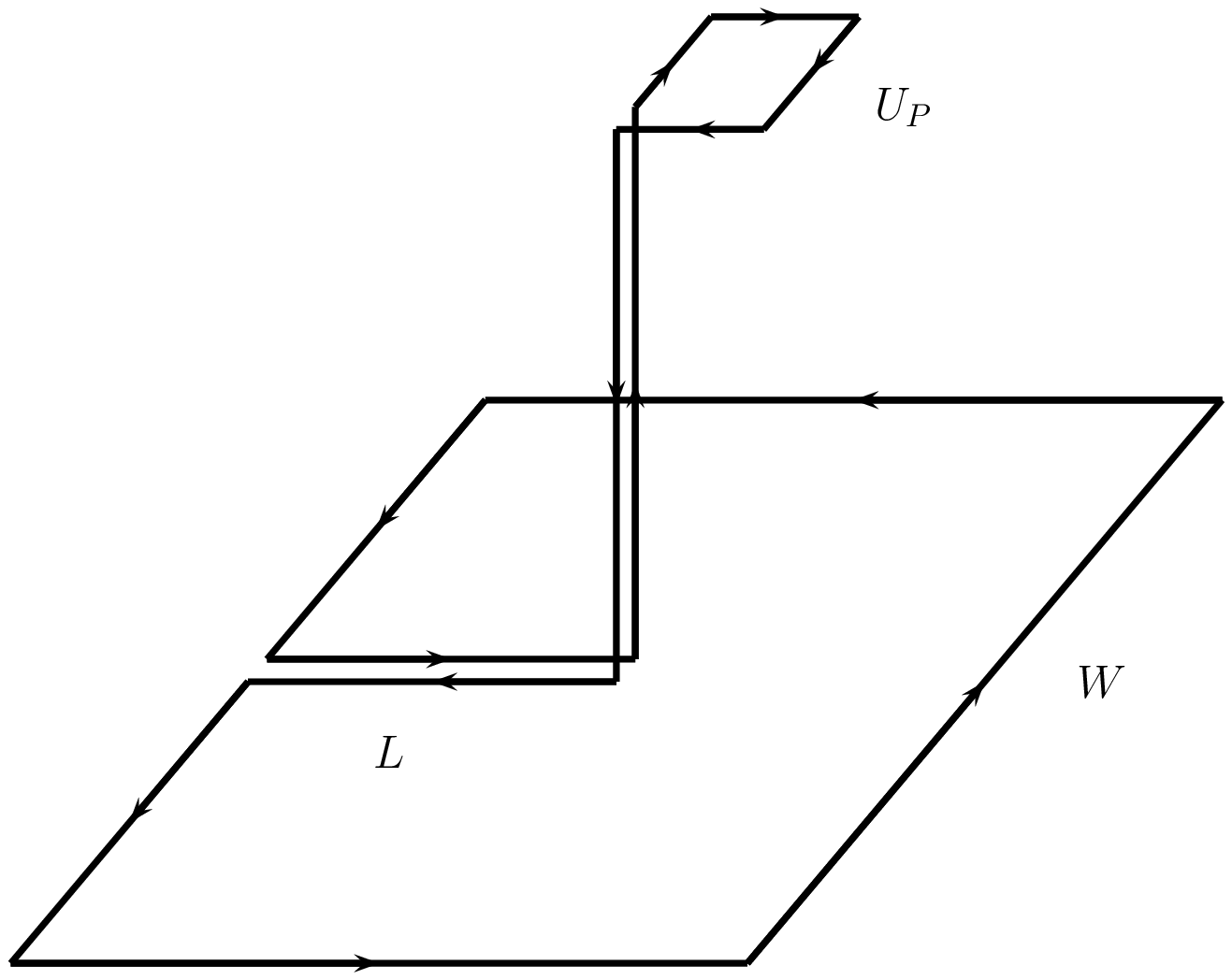} 
\hspace{0.5cm}
\includegraphics[width=0.45\textwidth]{Ex_vs_dist_5.90_cool10.eps} 
\caption[]{(Left) The connected correlator~(\ref{rhoW}) between the
plaquette $U_p$ and the Wilson loop. The subtraction appearing in the
definition of correlator is not explicitly drawn. (Right)
Longitudinal component of the chromoelectric field versus the 
distance $x_t$ at $\beta=5.9$ after 10 cooling steps.}
\label{Fig:correlator}
\end{figure}

We use a connected correlator (Fig.~\ref{Fig:correlator})(left) to explore the 
field configurations produced by a static quark-antiquark pair ($N$ is the 
number of colors)~\cite{DiGiacomo:1989yp,DiGiacomo:1990hc,Kuzmenko:2000bq,
DiGiacomo:2000va}:
\begin{equation}
\label{rhoW}
\rho_W = \frac{\left\langle {\rm tr}
\left( W L U_P L^{\dagger} \right)  \right\rangle}
              { \left\langle {\rm tr} (W) \right\rangle }
 - \frac{1}{N} \,
\frac{\left\langle {\rm tr} (U_P) {\rm tr} (W)  \right\rangle}
              { \left\langle {\rm tr} (W) \right\rangle } \;.
\end{equation}
In the naive continuum limit~\cite{DiGiacomo:1990hc}
\begin{equation}
\label{rhoWlimcont}
\rho_W  \stackrel{a \rightarrow 0}{\longrightarrow} a^2 g \left[ \left\langle
F_{\mu\nu}\right\rangle_{q\bar{q}} - \left\langle F_{\mu\nu}
\right\rangle_0 \right] \;,
\end{equation}
so that
\begin{equation}
\label{fieldstrength}
\hat F_{\mu\nu}(x) = \sqrt\frac{\beta}{2 N} \, \rho_W(x)
\end{equation}
The configuration with plaquette parallel to the Wilson loop leads to
chromoelectric field longitudinal to the axis defined by the static quarks.

\subsection{The case of SU(2) gauge theory}

The case of the SU(2) gauge theory was studied 
in~\cite{Cea:1992sd,Cea:1992vx,Cea:1993pi,Cea:1994ed,Cea:1994aj,Cea:1995zt}.
The main result was that the flux tube is almost completely formed by the 
longitudinal chromoelectric field, $E_l$, which is constant along the flux 
and rapidly decreasing in the transverse direction $x_t$.
In the dual Meissner effect interpretation, the transverse shape 
of $E_l$ is the dual version of the Abrikosov vortex field 
distribution~\cite{Cea:1992sd,Cea:1992vx,Cea:1993pi,Cea:1994ed,
Cea:1994aj,Cea:1995zt} and therefore must obey
\begin{equation}
\label{London}
E_l(x_t) = \frac{\Phi}{2 \pi} \mu^2 K_0(\mu x_t) \;,\;\;\;\;\; x_t > 0 \;,
\end{equation}
where $\Phi$ is the external flux and $\lambda=1/\mu$ is the 
London penetration length (this is valid for $\lambda \gg \xi$, with $\xi$ 
the coherence length of a type-II superconductor). 

In the past numerical study (lattices $16^4$, $20^4$ and $24^4$, statistics 
20-100)~\cite{Cea:1995zt} approximate scaling was found,
$\mu/\sqrt{\sigma}=4.04(18)$ 
(i.e. $\lambda=0.118(5)$~fm for $\sqrt{\sigma}=420$~MeV). In this work
($20^4$ lattice, statistics $1000$), we find $\mu/\sqrt{\sigma}=4.21(16)$
-- see Ref.~\cite{noi} for details.

\subsection{The case of SU(3) gauge theory}

The main motivation for repeating the study in SU(3) is to verify the
scaling of $\mu$ with the string tension and to compare
the resulting determination of $\mu/\sqrt{\sigma}$ with SU(2).
This result should provide us with important reference values, that any 
approach aiming at explaining confinement should be able to accommodate.

We performed numerical simulations with the Wilson action and periodic
boundary conditions, using a the Cabibbo-Marinari algorithm~\cite{Cabibbo:1982zn}, 
combined with overrelaxation on SU(2) subgroups. The summary of 
$\beta$ values, lattice size, Wilson
loop size and statistics is given in Table~\ref{Table:runs}. The lattice
size $L$ has been chosen such that the combination $L\sqrt{\sigma}\gtrsim 4$.
The size of the Wilson loop entering the definition of the operator given
in Eq.~(\ref{rhoW}) has been fixed at $L/2-2a$. In order to reduce the 
autocorrelation time, measurements were taken after 10 updatings. The 
error analysis was performed by the jackknife method over bins at different 
blocking levels.

\begin{table}[tb]
\begin{center}
\begin{tabular}{|c|c|c|c|}
\hline
$\beta$ & lattice & Wilson loop & statistics \\
\hline
 5.90   & $18^4$  & $7\times 7$  &  5.k   \\ 
 6.00   & $20^4$  & $8\times 8$  &  4.5k \\ 
 6.05   & $22^4$  & $9\times 9$  &  3.6k \\ 
 6.10   & $24^4$  &$10\times 10$ &  2.4k \\ 
\hline
\end{tabular}
\end{center}
\caption[]{Summary of the Monte Carlo simulations.}
\label{Table:runs}
\end{table}

In order to reduce the quantum fluctuations we adopted the controlled
cooling algorithm. It is known~\cite{Campostrini:1989ts} that 
by cooling in a smooth way equilibrium configurations, quantum fluctuations 
are reduced by a few order of magnitude, while the string tension survives 
and shows a plateau. We shall show below that the penetration length behaves 
in a similar way. The details of the cooling procedure are described in 
Ref.~\cite{Cea:1995zt} for the case of SU(2). Here we adapted the procedure
to the case of SU(3), by applying successively this algorithm to various 
SU(2) subgroups. The control parameter $\delta$ was fixed at the value 
0.0354, as in Ref.~\cite{Cea:1995zt}.

A novelty with respect to the study of Ref.~\cite{Cea:1995zt}
is related with the construction of the lattice operator given in
Eq.~(\ref{rhoW}). If the Wilson loop lies on the plane, say, 1-2, then
the Schwinger line can leave the plane 1-2 in the direction, say, 3;
before attaching the plaquette to the Schwinger line, the latter can be 
prolongated further in the direction 4, by one or two links. In this way,
by varying the length of the Schwinger line in the direction 3, one can 
obtain a large set of distances $x_t/a$ between the center of the plaquette 
and the center of the Wilson loop, both integer and non-integer. On each 
configuration we averaged over all possible directions for the relative 
orientation of the Wilson loop to the Schwinger line.

The general strategy underlying this work is the following:
(1) for each $\beta$ we generate an ensemble of thermalized configurations
and, correspondingly, ensembles of ``cooled'' configurations after a number
of cooling steps ranging from 5 to 16;
(2) for different values of the distance $x_t$, the longitudinal component 
of the chromoelectric field, averaged over each cooled ensemble of 
configurations, is then determined by means of the operator~(\ref{rhoW}),
with the help of Eq.~(\ref{fieldstrength}) (see, for example, 
Fig.~\ref{Fig:correlator}(right), which shows $E_l(x_t)$ averaged over the 
ensemble at $\beta=5.90$ after 10 cooling steps);
(3) for each cooling step, data for $E_l(x_t)$ are fitted with the function
given in Eq.~(\ref{London}) and the parameters $\mu$ and $\Phi$ are extracted;
(4) a plateau is then searched in the plot for $\mu$ and $\Phi$ versus
the cooling step.

In Table~\ref{Table:mu_fit} we report the results for $a\mu$ of the fit at 
the four $\beta$ values considered in this work for one selected cooling step. 
The table with the results for the other fit parameter, $\Phi$, can be found 
in Ref.~\cite{noi}. When the fit is done on all available data 
for $E_x(x_t)$, above a certain $x_{t,{\rm{min}}}$, the $\chi^2$/d.o.f. is 
very high, thus reflecting the wiggling of data due to the inclusion of 
non-integer distances $x_t/a$. When the fit is restricted to integer values 
of $x_t/a$, the $\chi^2$/d.o.f. turns out to be very reasonable. Remarkably, 
the resulting parameters obtained with the two fitting procedures agree very 
well.

\begin{figure}[tb]
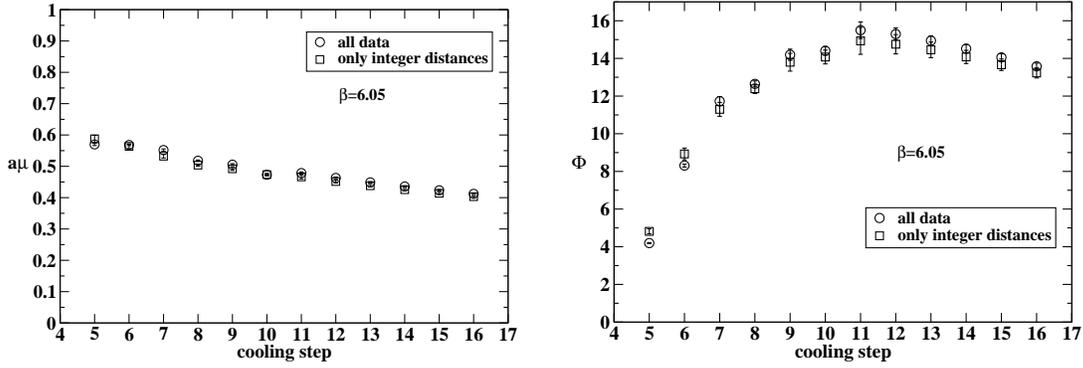

\centering
\includegraphics[width=0.45\textwidth]{mu_vs_cooling_6.05.eps} 
\hspace{0.5cm}
\includegraphics[width=0.45\textwidth]{phi_vs_cooling_6.05.eps} 
\caption[]{(Left) The inverse of the penetration length $a\mu$ at $\beta=6.05$
versus the cooling step. Data are obtained by fitting the transverse profile 
of the longitudinal chromoelectric field with the function~(\ref{London}); 
circles correspond to fit to all available data of $E_l(x_t)$ starting from 
a certain $x_{t,{\rm min}}$, while squares correspond to fit of $E_l(x_t)$ 
for integer values of $x_t/a$.
(Right) The same for the amplitude of the longitudinal chromoelectric field $\Phi$.}
\label{Fig:mu_phi_vs_cooling_6.05} 
\end{figure}

In Figs.~\ref{Fig:mu_phi_vs_cooling_6.05} we show the behavior of $a\mu$ and 
$\Phi$ with the cooling step at $\beta=6.05$. Similar figures for the other 
values of $\beta$ are given in Ref.~\cite{noi}. A short plateau is visible, 
except for the case of $\mu$ at $\beta=5.90$. We take as ``plateau'' value 
for $\mu$ the value corresponding to the number of cooling steps given in the 
second column of Table~\ref{Table:mu_fit}.

\begin{table}[tb]
\tabcolsep .2cm
\begin{center}
\begin{tabular}{|c|c|c|c|c|c|}
\hline
$\beta$ & cooling  step & $a\mu$ & $\chi^2$/d.o.f. & $x_{t,\rm{min}}/a$ & data 
set \\
\hline
 5.90   & 10 & 0.5577(12) & 626. & 6 & all data \\
 6.00   &  9 & 0.51015(92)& 383. & 6 & all data \\
 6.05   & 10 & 0.4730(13) & 133. & 7 & all data \\
 6.10   & 10 & 0.4357(20) &  27. & 7 & all data \\
\hline
 5.90   & 10 & 0.5557(40) & 1.22 & 7 & integer $x_t/a$ \\
 6.00   &  9 & 0.5099(28) & 2.56 & 9 & integer $x_t/a$ \\
 6.05   & 10 & 0.4735(39) & 1.08 & 8 & integer $x_t/a$ \\
 6.10   & 10 & 0.4349(56) & 0.25 & 8 & integer $x_t/a$ \\
\hline
\end{tabular}
\end{center}
\caption[]{Summary of the fit values for $a\mu$.} 
\label{Table:mu_fit}
\end{table}

Finally, we studied the scaling of the ``plateau'' values of $a\mu$ 
with the string tension. For this purpose, we have expressed these 
values of $a\mu$ in units of $\sqrt\sigma$, using the parameterization 
\beq
\label{sigma}
a\sqrt{\sigma}(g)=f_{SU(3)}(g^2)[1+0.2731\,\hat{a}^2(g)
-0.01545\,\hat{a}^4(g) +0.01975\,\hat{a}^6(g)]/0.01364 \;, 
\eeq
\[
\hat{a}(g) = \frac{f_{SU(3)}(g^2)}{f_{SU(3)}(g^2(\beta=6))} \;, \;\;\;
\beta=\frac{6}{g^2} \,, \;\;\; 5.6 \leq \beta \leq 6.5\;,
\]
\beq
\label{fsun}
f_{SU(3)}(g^2) = \left( {b_0 g^2}\right)^{- b_1/2b_0^2} 
\, \exp \left( - \frac{1}{2 b_0 g^2}\right) \,, \;\;\;\;\;
b_0=\frac{11}{(4\pi)^2}\;, \;\;\; b_1=\frac{102}{(4\pi)^4}\;,
\eeq
given in Ref.~\cite{Edwards:1998xf}.

\begin{figure}[tb]
\centering
\includegraphics[width=0.45\textwidth]{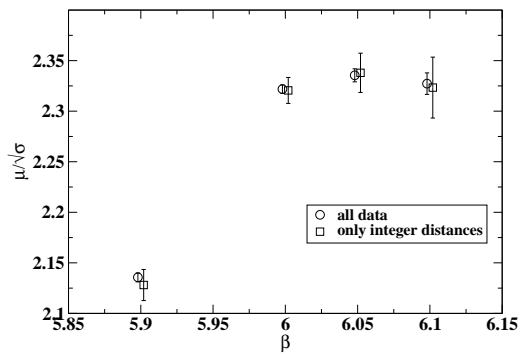} 
\caption[]{Scaling of the inverse London penetration length with $\sqrt\sigma$
versus $\beta$. Data have been slightly shifted on the horizontal axis for 
the sake of readability.}
\label{Fig:scaling} 
\end{figure}

Figure~\ref{Fig:scaling} suggests that the ratio $\mu/\sqrt\sigma$ displays 
a nice plateau in $\beta$, as soon as $\beta$ is larger than 6.
The scaling of $\mu$ is a natural consequence of the fact that the penetration
length is a physical quantity related to the size $D$ of the flux
tube~\cite{Cea:1992sd,Cea:1992vx}, $D  \simeq  2/\mu$.
We get as estimate for the penetration length in SU(3) gauge theory,
$\mu/\sqrt{\sigma} = 2.325 (5)$, corresponding to $\mu$ =  0.977(2) GeV.
We observe that this value is in nice agreement with the determinations
of Ref.~\cite{Bicudo:2010gv}, obtained by using  correlators of plaquette and 
Wilson loops not connected by the Schwinger line, thus 
leading to the (more noisy) squared chromoelectric and chromomagnetic 
fields.

We note that the ratio between the penetration lengths for the SU(2) gauge 
theory and the SU(3) gauge theory is $\mu_{\text{SU(2)}}/\mu_{\text{SU(3)}}$ 
= 1.81(7).
This result recalls analogous behavior seen in a different study of SU(2) and 
SU(3) vacuum in a constant external chromomagnetic background 
field~\cite{Cea:2005td}. In Ref.~\cite{Cea:2005td} 
numerical evidence that the deconfinement temperature for SU(2) and SU(3) 
gauge systems in a constant Abelian chromomagnetic field decreases when the 
strength of the applied field increases was given. 
Moreover, as discussed in Refs.~\cite{Cea:2001an,Cea:2005td,Cea:2007yv}, 
above a critical strength  $\sqrt{gH_c}$ of the chromomagnetic external 
background field the deconfined phase extends to very low temperatures. It 
was found~\cite{Cea:2005td} 
that the ratio between the critical field strengths for SU(2) and SU(3) gauge 
theories is $ \sqrt{gH_c}|_{\text{SU(2)}}/\sqrt{gH_c}|_{\text{SU(3)}}$ 
= 2.03(17), in remarkable agreement with the ratio between the penetration 
lengths for SU(2) and SU(3).
As stressed in the Conclusions of Ref.~\cite{Cea:2005td}, the peculiar 
dependence of the deconfinement temperature on the strength of the Abelian 
chromomagnetic field $gH$ could be naturally explained if the vacuum behaved 
as a disordered chromomagnetic condensate which confines color charges due 
both to the presence of a mass gap and the absence
of color long range order, such as in the Feynman picture for Yang-Mills 
theory in  (2+1) dimensions~\cite{Feynman:1981ss}.
The circumstance that  ratio between the SU(2) and SU(3) penetration lengths 
agrees within errors with the above discussed ratio of the critical 
chromomagnetic fields, suggests  us that the Feynman picture of the 
Yang-Mills vacuum could be a useful guide to understand the dynamics of color 
confinement.


\begin{thebibliography}{44}

\bibitem{Bander:1980mu}
M.~Bander, Phys. Rept. \textbf{75}, 205 (1981).

\bibitem{Greensite:2003bk}
J.~Greensite, Prog. Part. Nucl. Phys. \textbf{51}, 1 (2003).

\bibitem{Fukugita:1983du}
M.~Fukugita and T.~Niuya, Phys. Lett. \textbf{B132}, 374 (1983).

\bibitem{Kiskis:1984ru}
J.E.~Kiskis and K.~Sparks, Phys. Rev. \textbf{D30},  1326 (1984).

\bibitem{Flower:1985gs}
J.W. Flower and S.W. Otto, Phys. Lett. \textbf{B160}, 128 (1985).

\bibitem{Wosiek:1987kx}
J.~Wosiek and R.W. Haymaker, Phys. Rev. \textbf{D36}, 3297 (1987).

\bibitem{DiGiacomo:1989yp}
A.~Di~Giacomo {\it et al.}, Phys. Lett. \textbf{B236}, 199 (1990).

\bibitem{DiGiacomo:1990hc}
A.~Di~Giacomo {\it et al.}, Nucl. Phys. \textbf{B347}, 441 (1990).

\bibitem{Singh:1993jj}
V.~Singh {\it et al}, Phys. Lett. \textbf{B306}, 115 (1993).

\bibitem{Cea:1992sd}
P.~Cea and L. Cosmai, Nucl. Phys. Proc. Suppl. \textbf{{30}}, 572 (1993).

\bibitem{Matsubara:1993nq}
Y.~Matsubara {\it et al.}, Nucl. Phys. Proc. Suppl. \textbf{34},
176 (1994).

\bibitem{Cea:1992vx}
P.~Cea and L.~Cosmai, Nuovo Cim. \textbf{A107}, 541 (1994).

\bibitem{Cea:1993pi}
P.~Cea and L.~Cosmai, Nucl. Phys. Proc. Suppl. \textbf{34}, 219 (1994).

\bibitem{Cea:1994ed}
P.~Cea and L.~Cosmai, Phys. Lett. \textbf{B349}, 343 (1995).

\bibitem{Cea:1994aj}
P.~Cea and L.~Cosmai, Nucl. Phys. Proc. Suppl. \textbf{42}, 225 (1995).

\bibitem{Cea:1995zt}
P.~Cea and L.~Cosmai, Phys. Rev. \textbf{D52}, 5152 (1995).

\bibitem{Bali:1994de}
G.S. Bali {\it et al.}, Phys. Rev. \textbf{D51}, 5165 (1995).

\bibitem{Haymaker:2005py}
R.W. Haymaker and T.~Matsuki, Phys. Rev. \textbf{D75}, 014501 (2007).

\bibitem{D'Alessandro:2006ug}
A.~D'Alessandro {\it et al.}, Nucl. Phys. \textbf{B774}, 168 (2007).

\bibitem{'tHooft:1976ep}
G.~'t~Hooft, in
\emph{High Energy Physics, EPS International Conference, Palermo, 1975}, 
edited by A.~Zichichi (1975).

\bibitem{Mandelstam:1974pi}
S.~Mandelstam, Phys. Rept. \textbf{23}, 245 (1976).

\bibitem{Ripka:2003vv}
G.~Ripka, hep-ph/0310102.

\bibitem{Abrikosov:1957aa}
A.A. Abrikosov, Soviet Physics JETP \textbf{5}, 1174 (1957).

\bibitem{Shiba:1994db}
H.~Shiba and T.~Suzuki, Phys. Lett. \textbf{B351}, 519 (1995).

\bibitem{Arasaki:1996sm}
N.~Arasaki {\it et al.}, Phys. Lett. \textbf{B395}, 275 (1997). 

\bibitem{Cea:2000zr}
P.~Cea and L.~Cosmai, Phys. Rev. \textbf{D62}, 094510 (2000).

\bibitem{Cea:2001an}
P.~Cea and L.~Cosmai, JHEP \textbf{{11}}, 064 (2001).

\bibitem{DiGiacomo:1999fa}
A.~Di~Giacomo {\it et al.}, Phys. Rev. \textbf{D61}, 034503 (2000).

\bibitem{DiGiacomo:1999fb}
A.~Di~Giacomo {\it et al.}, Phys. Rev. \textbf{D61}, 034504 (2000).

\bibitem{Carmona:2001ja}
J.M. Carmona {\it et al.}, Phys. Rev. \textbf{D64}, 114507 (2001).

\bibitem{Cea:2004ux}
P.~Cea {\it et al.}, {JHEP} \textbf{{02}}, 018 (2004).

\bibitem{D'Alessandro:2010xg}
A.~D'Alessandro {\it et al.}, Phys. Rev. \textbf{D81}, 094501 (2010).

\bibitem{'tHooft:2004th}
G.~'t~Hooft, hep-th/0408183.

\bibitem{Kuzmenko:2000bq}
D.S. Kuzmenko and Y.A. Simonov, Phys. Lett. \textbf{B494}, 81 (2000).

\bibitem{DiGiacomo:2000va}
A.~Di~Giacomo {\it et al.}, Phys. Rept. \textbf{372}, 319 (2002).

\bibitem{noi}
M.S.~Cardaci {\it et al.}, Phys. Rev. \textbf{D83}, 014502 (2011).

\bibitem{Cabibbo:1982zn}
N.~Cabibbo and E.~Marinari, Phys. Lett. \textbf{B119}, 387 (1982).

\bibitem{Campostrini:1989ts}
M.~Campostrini {\it et al.}, Phys. Lett. \textbf{B225}, 403 (1989).

\bibitem{Edwards:1998xf}
R.G. Edwards {\it et al.}, Nucl. Phys. \textbf{B517}, 377 (1998).

\bibitem{Bicudo:2010gv}
P.~Bicudo {\it et al.}, PoS {\bf LATTICE2010}, 268 (2010).

\bibitem{Cea:2005td}
P.~Cea and L.~Cosmai, JHEP \textbf{08}, 079 (2005).

\bibitem{Cea:2007yv}
P.~Cea {\it et al.}, JHEP \textbf{12}, 097 (2007).

\bibitem{Feynman:1981ss}
R.P. Feynman, Nucl. Phys. \textbf{B188}, 479 (1981).

\end{thebibliography}
\end{document}